\documentclass[aps,prl,twocolumn,amsmath,amssymb,showpacs,superscriptaddress]{revtex4-1}
\usepackage{graphicx}
\usepackage{dcolumn}
\usepackage[mathlines]{lineno}
\usepackage{hyperref}
\usepackage{bm}
\usepackage{epstopdf}
\usepackage{lipsum}
\usepackage{epsfig}
\usepackage{bbold}
\usepackage{graphicx}

\newcommand{\ket}[1]{\ensuremath{\left|#1\right\rangle}}

\usepackage{color}

\begin{document}
	\title{A witness to quantify high-dimensional entanglement}
	\author{Bienvenu~Ndagano}
	\email[Corresponding author: ]{nibienvenu@gmail.com}
	\affiliation{School of Physics, University of the Witwatersrand, Private Bag 3, Wits 2050, South Africa}
	\author{Andrew~Forbes}
	\affiliation{School of Physics, University of the Witwatersrand, Private Bag 3, Wits 2050, South Africa}
	
	\date{\today}
	
	\begin{abstract}
\noindent Exploiting photonic high-dimensional entanglement allows one to expand the bandwidth of quantum communication and information processing protocols through increased photon information capacity. However, the characterization of entanglement in higher dimensions is experimentally challenging because the number of measurements required scales unfavourably with dimension.  While bounds can be used to certify high-dimensional entanglement, they do not quantify the degree to which quantum states are entangled. Here, we propose a quantitative measure that is both an entanglement measure and a dimension witness, which we coin the ${\cal P}$-concurrence. We derive this measure by requiring entanglement to extend to qubit subspaces that constitute the high-dimensional state. The computation of the ${\cal P}$-concurrence is not contingent on reconstructing the full density matrix, and requires less measurements compared to standard quantum state tomography by orders of magnitude. This allows for faster and more efficient characterization of high-dimensional quantum states.
		
	\end{abstract}
	\maketitle
	

Quantum entanglement with photons is an enabling resource in quantum communication and information processing where, exploiting correlations between entangled photon pairs allows one to realise protocols unique to quantum mechanics, for example, quantum computing \cite{Knill2010,Ladd2010}, quantum teleportation \cite{Ma2012a,Yin2012} and quantum key distribution \cite{Ursin2007,Lo2014,Liao2017}. There is a growing interest in engineering and utilizing photonic states entangled in higher dimensions (higher than 2) \cite{Mair2001,Romero2012, Krenn2016,Erhard2017}. This has been nurtured by the promise of higher information capacity per photon and increased security \cite{Bechmann-Pasquinucci2000,Cerf2002, Scarani2009}.

Certification is a crucial step when engineering and utilizing high-dimensional quantum states. This can be done in two ways: quantum state reconstruction or violation of entanglement bounds. Quantum state tomography (QST) is the tool of choice when reconstructing a quantum state. It involves the measurement of a set of observables from which the density matrix can be estimated via numerical optimization \cite{Jack2009}. One then computes the fidelity to a given maximally entangled state to certify high-dimensional entanglement. Through QST, entanglement has been certified in up to dimension $d = 8$ \cite{Agnew2011}. However, QST is computationally expensive as the number of measurement scales with ${\cal O}(d^4)$. Agnew \textit{et al.} highlight in \cite{Agnew2011} that the measurement process takes over 40 h for $d=8$ with an integration time of 10 s.

Alternatively one can compute the degree of entanglement. The entanglement of formation adequately bounds the degree of entanglement and is simple to compute in arbitrary dimensions for pure states. The concurrence is another measure of entanglement that has been shown theoretically \cite{Hill1997,Wootters1998,Wootters2001}, and used experimentally \cite{McLaren2014a, Romero2012, Zhang2017, Ndagano2017}, to characterize entanglement at the qubit level, but not for high-dimensional states. We also note that the negativity \cite{Vidal2002} has been used, experimentally, to characterize both qubit and qutrit entanglement \cite{Jack2009,Zhang2016}. The inconvenience with these measures of entanglement is that for non-maximally entangled states they do not in general account for the dimensionality of the system. 

Rather than reconstructing the quantum state, one has the option of using entanglement bounds to certify high-dimensional entanglement. The Bell inequality is a common entanglement bound. Generalized to arbitrary dimensions by Collins \textit{et al.} \cite{Collins2002}, it has been used to certify entanglement in up to dimension $d=11$ \cite{Yin2017,Vaziri2002, Groblacher2006,Dada2011a}. Using a predetermined measurement setting, one computes the $d$-dimensional Bell parameter $S_d$ which is bounded above by $S_d = 2$ for any theory based on local realism (classical limit). Thus, violation of this bound in a multi-dimensional setting is enough to certify high-dimensional entanglement. Similarly, one could determine the Schmidt number associated with a particular \textit{d}-dimensional state and show that it exceeds both the classical limit (1 for separable states) and the bound allowed for entangled states in $d-1$ dimensions \cite{Terhal2000, Sanpera2001, Guhne2009, Bavaresco2017}.  

\begin{figure}[ht]
	\begin{center}
		\includegraphics[width = \linewidth]{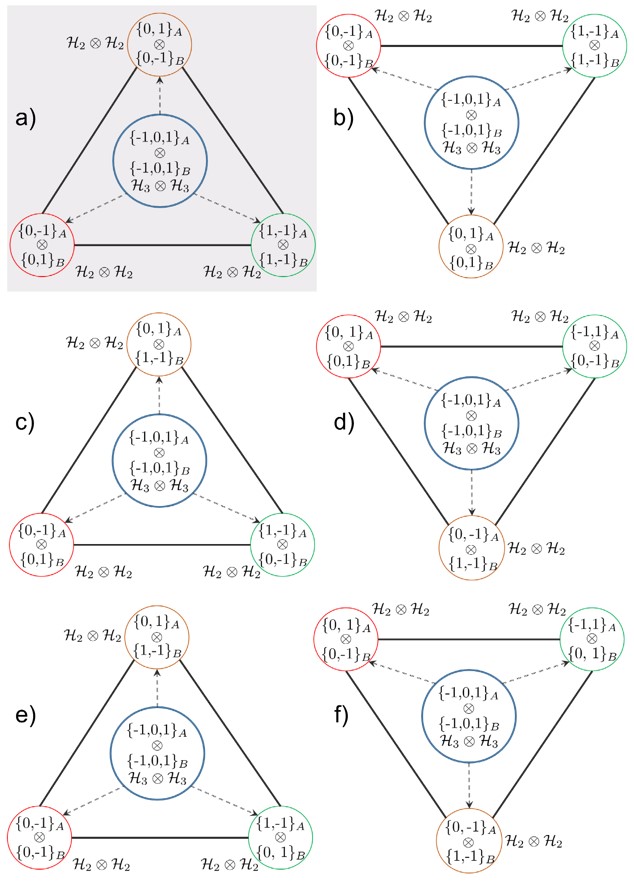}
	\end{center}
	\caption{Graphical partition of high-dimensional entanglement. The state space ${\cal H}_3 \otimes {\cal H}_3$ of a bi-photon qutrit is partitioned in two-dimensional subspaces ${\cal H}_2 \otimes {\cal H}_2$ shown on the vertices of the triangular pattern. (a)-(f) Show the 6 distinct partitions of a qutrit state space into qubit subspaces}   
	\label{fig:Figure1}
\end{figure}

Here we introduce a new measure that acts both as a dimension witness and as an estimation of the degree of entanglement.  As discussed previously, a \textit{d}-dimensional entangled system may violate the entanglement bound in \textit{d-1} dimensions. The implication is that, for such a system, entanglement must also exist in the constitutive subspaces. The measure we present, the ${\cal P}$-concurrence, exploits this principle by expressing the degree of entanglement in high-dimension, as a function of the entanglement in the qubit subspaces. By requiring entanglement to exist in qubit subspaces, the ${\cal P}$-concurrence computed in \textit{d}-dimensions, is non-vanishing for any system exhibiting entanglement correlations in dimension $\geq d$, thus certifying entanglement dimensionality. Furthermore the ${\cal P}$-concurrence expresses the degree of entanglement as a value between 0 and 1. Lastly, we show that the number of measurements required to compute the ${\cal P}$-concurrence can be reduced significantly as compared to other approaches, for example, by one order of magnitude for $d=8$, leading to faster and more efficient characterization of high-dimensional entangled states.

Let us present the problem from an experimentalist's perspective. Alice, the experimentalist, wants to produce a qutrit state and determine its degree of entanglement. As is commonly the case, Alice uses a non-linear crystal to generate photon pairs via spontaneous parametric down-conversion (SPDC). Alice is interested in the following two-photon state
\begin{equation}
\ket{\Psi_{AB}} = {\cal N}\left(1\ket{0,0}+ \alpha\ket{1,-1} + \beta\ket{-1,1}\right), \label{eq:state1}
\end{equation}
where $\ket{\ell}$ are OAM eigenstates and ${\cal N}$ is a normalization constant: ${\cal N} = 1/\sqrt{1+\alpha^2+\beta^2}$. The probability amplitudes $\alpha$ and $\beta$, with $\alpha,\beta\in [0, 1]$, depend on experimental parameters. Here, we have assume that the phase information is carried by the kets. 

The state $\ket{\Psi_{AB}}$ can be projected on qubit subspaces ${\cal H}_2 \otimes {\cal H}_2$, spanned by the eigenstates shown at the vertices in Fig.~\ref{fig:Figure1} (a). Such a projection can be achieved with the appropriate operator ${\cal B}^k$, where $k$ denotes the distinct qubit spaces -- by distinct we mean that the set of qubit states of each photon is unique to each $k$ subspace. In the case of Alice's qutrit state, the qubit states produced after acting on $\ket{\Psi_{AB}}$ with ${\cal B}^k$ are as follows
\begin{eqnarray}
\ket{\psi^1_{AB}} &=& {\cal B}^1\ket{\Psi_{AB}} = {\cal N}_1 \left(1\ket{0,0} + \alpha\ket{1,-1}\right), \label{eq:substate1}\\
\ket{\psi^2_{AB}} &=& {\cal B}^2\ket{\Psi_{AB}} = {\cal N}_2 \left(1\ket{0,0} + \beta\ket{-1,1}\right), \label{eq:substate2}\\
\ket{\psi^3_{AB}} &=& {\cal B}^3\ket{\Psi_{AB}} = {\cal N}_3 \left(\alpha\ket{1,-1} + \beta\ket{-1,1}\right), \label{eq:substate3}
\end{eqnarray}
where ${\cal N}_k$ are normalization constants within each qubit space. Observe that projections of the high-dimensional state on qubit subspaces produces a set of entangled qubit states $\ket{\psi^k_{AB}}$ that each live in the subspaces at the vertices in Fig.~\ref{fig:Figure1}(a). We thus deduce the following property: \textit{A pure state, entangled in high-dimension, exhibits entanglement in all the constituent qubit subspaces that preserve the correlation between the entangled degrees of freedom.} The requirement of entanglement in the qubit subspaces is directly observed from Eqs. \ref{eq:substate1}-\ref{eq:substate3}. The additional requirement on correlation between degrees of freedom (DoFs) is implicit in Fig.~\ref{fig:Figure1}(a) and has a physical construct that will be addressed later.

We propose the following computation for the degree of entanglement in higher-dimensions: consider the two-dimensional spaces in Fig.~\ref{fig:Figure1}(a). Let ${\cal E}(\rho^k_2)$ be the degree of entanglement of the bi-photon qubit density matrix $\rho^k_2$ in the $k$th subspace, with $k = 1,2,...,K$. The number of $K$ distinct subspaces is given by 
\begin{equation}
K = \frac{d!}{(d-2)!2!}.
\end{equation}

For convenience, we will assume that $0\leq{\cal E}(\cdot)\leq1$. Using the language of statistics, we will substitute `degree of entanglement' by `probability of entanglement'. From the requirement that entanglement must exist in each of the subspaces, we express the `probability of entanglement' of the higher dimensional state $\rho$ as:
\begin{equation}
{\cal E}(\rho_{AB}) = {\cal E}(\rho^1_{AB})\cap{\cal E}(\rho^2_{AB})\cap...\cap{\cal E}(\rho^K_{AB}).
\end{equation}
Given that the subspaces are distinct, we treat the entanglement in each of the subspaces as an independent event, resulting in the following expression for the `probability of entanglement' in higher dimensions:
\begin{equation}
{\cal E}(\rho_{AB}) = {\cal E}(\rho^1_{AB})\times{\cal E}(\rho^2_{AB})\times...\times{\cal E}(\rho^K_{AB}).
\end{equation}
The concurrence, ${\cal C}$, is a well known continuous measure of entanglement for both pure and mixed states in two dimensions \cite{Hill1997,Wootters1998,Wootters2001}. We identified the concurrence as our measure of the entanglement to validate the above substitution to `probability of entanglement'; $0\leq {\cal C}(\rho)\leq 1$ where ${\cal C}(\rho) = 0$ means $\rho$ is separable and ${\cal C}(\rho) = 1$ means it is maximally entangled. Finally we arrive at the following measure of entanglement in high-dimension that we call the ${\cal P}$-concurrence 
\begin{equation}
{\cal P}(\rho_{AB}) = \prod_{k = 1}^{K}{\cal C}(\rho^k_{AB}),\label{eq:pconc}
\end{equation}
Note that the ${\cal P}$ is in reference to the probabilistic origin of the measure. Thus, the ${\cal P}$-concurrence estimates the degree of entanglement in $d$-dimensions: it is 0 if the state does not exhibits $d$-dimensional entanglement, and is 1 for $d$-dimensional maximally entangled states. 

\begin{figure}[t]
	\begin{center}
		\includegraphics[width = 0.9\linewidth]{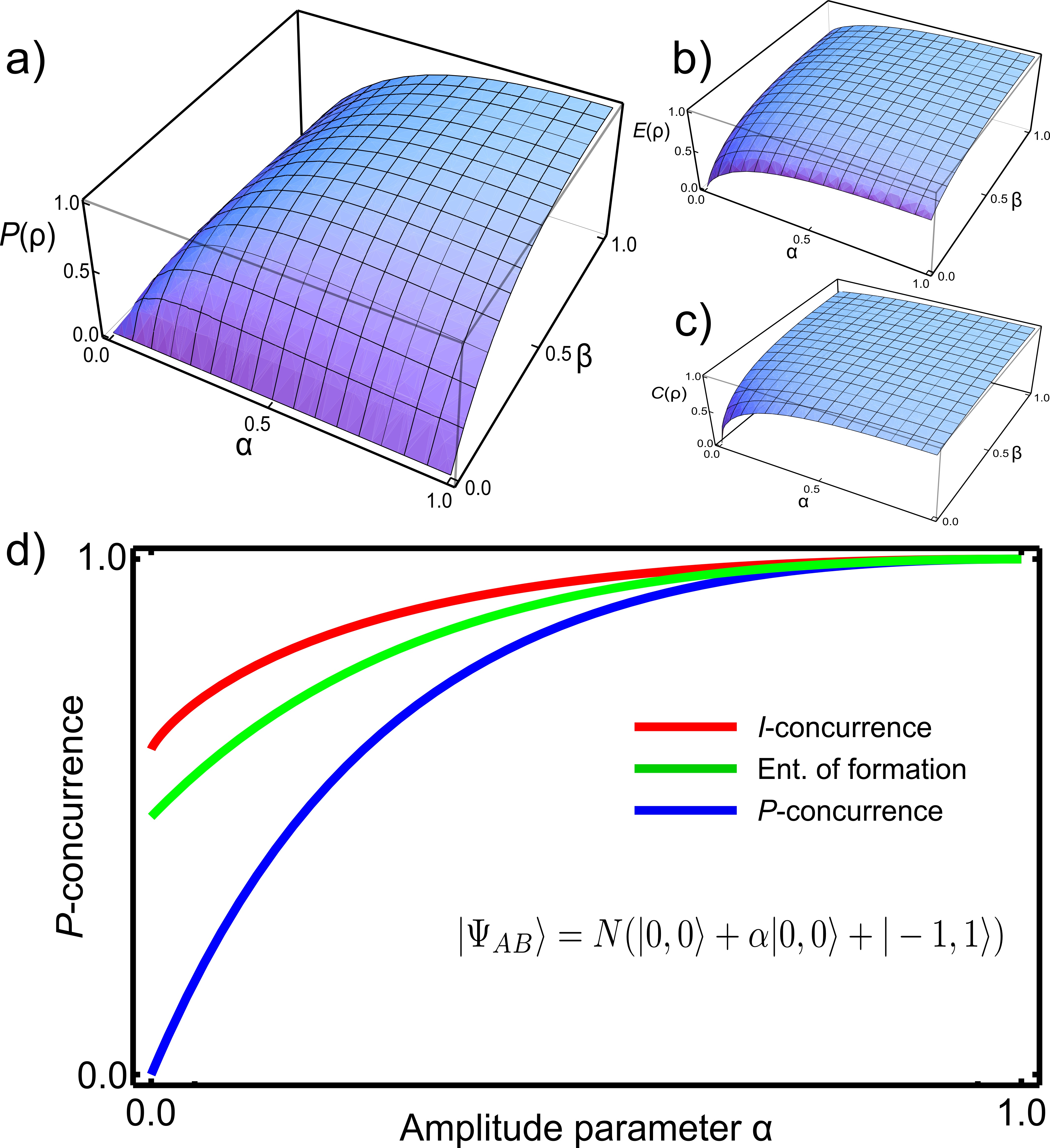}
	\end{center}
	\caption{Theoretical plot of (a) the ${\cal P}$-concurrence, (b) the entanglement of formation  and (c) the \textit{I}-concurrence of the qutrit state in Eq.~\ref{eq:state1}, as a function of the amplitude parameters $\alpha$ and $\beta$. (d) Evolution of entanglement measures from maximally entangled qubit to maximally entangled qutrit state.}   
	\label{fig:Figure3}
\end{figure}

The dimension witness aspect of the ${\cal P}$-concurrence is graphically depicted in Fig.~\ref{fig:Figure3}(a). Recall the parametric qutrit expression in Eq.~\ref{eq:state1} in terms of the real amplitudes $\alpha$ and $\beta$. We require a binary response to the question of whether or not Alice's bi-photon state is entangled in $d$-dimensions. In the above, we have shown that entanglement in $d$-dimensions is preserved down to qubit subspaces. With respect to the qutrit state in Eq.~\ref{eq:state1}, observe that for $\alpha, \beta$ = 0, $\ket{\Psi_{AB}}$ reduces to a qubit state. Furthermore, should $\alpha$ (or $\beta$) vanish, the concurrence in at least one of the $K$ qubit subspaces will vanish, resulting in a vanishing ${\cal P}$-concurrence according to Eq.~\ref{eq:pconc}. We conclude: \textit{a pure state is entangled in $d$-dimensions if and only if the ${\cal P}$-concurrence is non-zero}. This requirement is graphically depicted in Fig.~\ref{fig:Figure3}(a), where the computed ${\cal P}$-concurrence vanishes for qubit states; that is $\alpha, \beta = 0.$

The behaviour of the ${\cal P}$-concurrence is in contrast with other measures of entanglement. As a comparison, we consider the entanglement of formation and the $I$-concurrence. The entanglement of formation was first introduced by Bennett \textit{et al.} for mixed states \cite{Bennett1996}, and is defined as follows
\begin{equation}
E(\rho_{AB}) = -\text{Tr}\left[\rho_{A}\log_2(\rho_{A})\right] = -\text{Tr}\left[\rho_{B}\log_2(\rho_{B})\right], \label{eq:EOF}
\end{equation}
where $\rho_{A}$ ($\rho_{B}$) is the partial trace over subsystem $A$ ($B$). The $I$-concurrence is a generalization of the qubit concurrence to arbitrary dimensions. It was introduced by Rungta \textit{et al.} \cite{Rungta2001}, and takes the following form
\begin{equation}
C = \sqrt{2\left[1-\text{Tr}(\rho_{A}^2)\right]}. \label{eq:Conc}
\end{equation} 
Note that both the entanglement of formation and the \textit{I}-concurrence vanish only for separable states; that is $\alpha = \beta = 0$, as shown in Figs.~\ref{fig:Figure3}(b) and (c) -- both measures were normalized in three-dimensions. Observe that contrary to the ${\cal P}$-concurrence in Fig.~\ref{fig:Figure3}(a), Neither $E$ nor $C$ can distinguish qutrits from qubit states in general; that is because, there exists a set of qutrit states ($\alpha,\beta \neq 0$) with equal, and even lower degree of entanglement compared to qubits states ($\alpha,\beta = 0$). The dimension dependent aspect of the ${\cal P}$-concurrence is further demonstrated in Fig.~\ref{fig:Figure3}(d), where we compare all three measures of entanglement for a qutrit state shown in the inset. For $\alpha = 0$, the state $\ket{\Psi_{AB}}$ is not entangled in three-dimensions, hence the three-dimensional ${\cal P}$-concurrence vanishes, unlike the other two measures that are non-zero. However, they are all maximum for a maximally entangled qutrit state $(\alpha = 1)$.

\begin{figure*}[ht]
	\begin{center}
		\includegraphics[width = 0.9\linewidth]{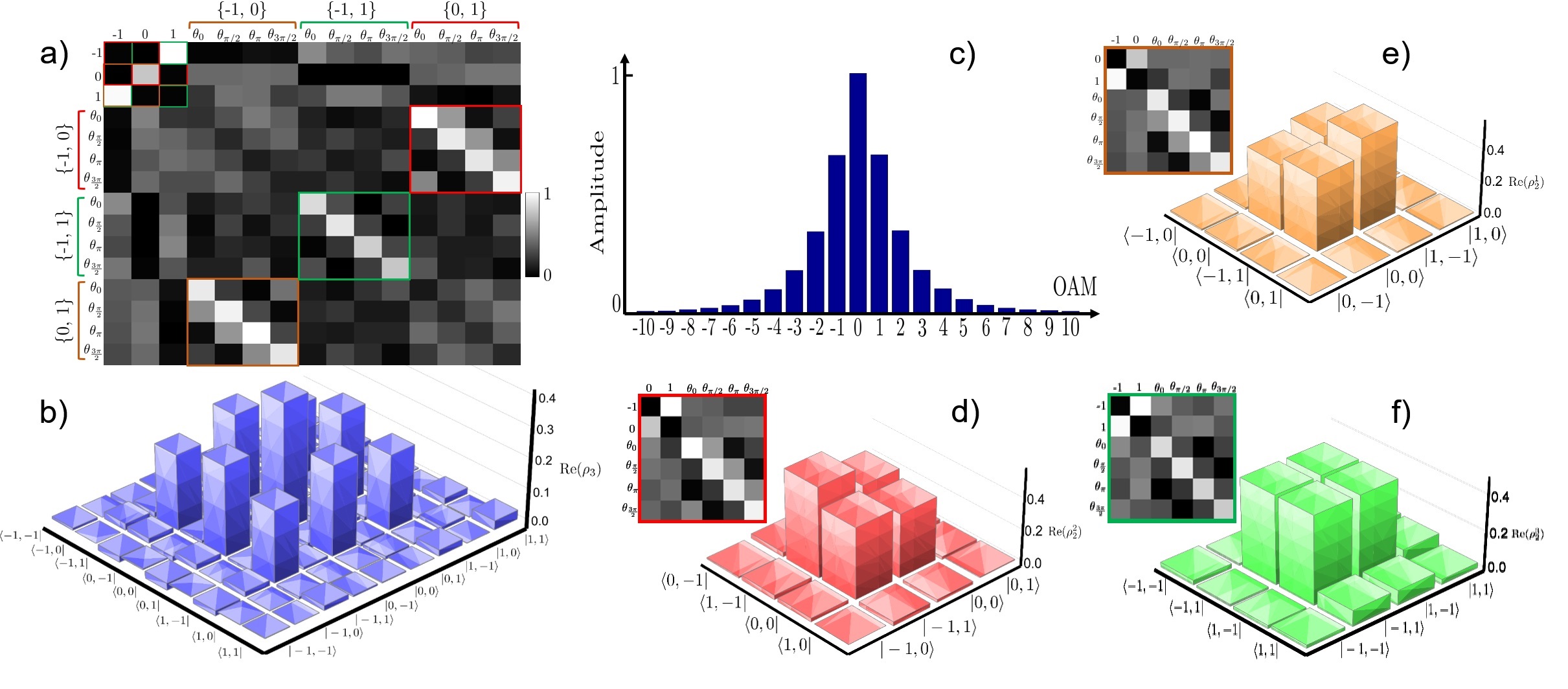}
	\end{center}
	\caption{(a) Alice performs a state tomography of her qutrit state by projecting photon A and photon B onto pure OAM states $\ket{\ell}$, as well as superposition states $\ket{\theta} = (\ket{\ell} + \exp(i\theta\ket{-\ell})/\sqrt{2}$ that label the rows and columns. (b) From the state tomography, Alice reconstructs the the high-dimensional density matrix. (c) Shows the OAM spectrum produces by Alice's entanglement source. (d)-(f) From the high-dimensional state tomography, Alice can extracts tomography data of the qubit subspaces and reconstruct the corresponding density matrices to compute the ${\cal P}$-concurrence.}   
	\label{fig:Figure2}
\end{figure*}

We use our measure of entanglement to characterize experimentally a pair of photons, entangled in orbital angular momentum (OAM). Alice performs a state tomography with mutually unbiased bases (MUBs) as shown in Fig.~\ref{fig:Figure2}(a) (a detailed description of the high-dimensional state tomography is presented in \cite{Agnew2011}) and reconstructs the high-dimensional density matrix shown in Fig.~\ref{fig:Figure2}(b). Note that the bi-photon state is not maximally entangled; this is supported by the measured OAM spectrum in Fig.~\ref{fig:Figure2}(c) that shows decaying probability amplitudes with increasing OAM. From the high-dimensional tomography, Alice extracts the qubit tomographies show in Figs.~\ref{fig:Figure2}(d)-(f), corresponding to that of the states in Eqs.~\ref{eq:substate1}-\ref{eq:substate3}, respectively.  For each of the subspaces, Alice reconstructs the qubit density matrices and computes the concurrence and fidelity values shown in Table~{\ref{table:1}}. The ${\cal P}$-concurrence of Alice's qutrit state is 0.80.
\begin{table}[h]
	\caption{Experimental concurrence and fidelity calculated for density matrices the two-dimensional subspaces of $\rho_{AB}$.}
	\begin{tabular}{|c|c|c|}
		\hline 
		Subspace $m$ & Concurrence & Fidelity \\ 
		\hline 
		$\{0,1\}_A\otimes\{0,-1\}_B$ & 0.92 & 0.95 \\ 
		\hline 
		$\{0,-1\}_A\otimes\{0,1\}_B$ & 0.93 & 0.96 \\ 
		\hline 
		$\{1,-1\}_A\otimes\{-1,1\}_B$ & 0.93 & 0.96 \\ 
		\hline
		\hline
		qutrit state  & \multicolumn{2}{c|}{${\cal P}$-concurrence}  \\ 
		\hline
		$\{1,0,-1\}_A\otimes\{-1,0,1\}_B$ & \multicolumn{2}{c|}{0.80}   \\ 
		\hline 
	\end{tabular}
	\label{table:1}
\end{table}
We highlight that the computation of the ${\cal P}$-concurrence is not contingent on reconstruction of the full high-dimensional state and requires less measurements. For example, to characterize the SPDC state studied by Agnew \textit{et al.} in \cite{Agnew2011} with $d=8$, the ${\cal P}$-concurrence requires $36\times K = 1\ 008$ measurements -- 36 is the number of measurements for an over-complete bi-photon qubit tomography.  This represents a measurement time of 2.8 hours, a significant reduction from the original 14 400 measurements over 40 hours. Also, computing the  ${\cal P}$-concurrence through reconstruction of qubit subspaces has the practical advantage that the number of MUBs is not a limiting factor, unlike in high-dimensional state tomographies with MUBs which can only be conducted in dimensions that are prime or powers of prime \cite{Wootters1989}.  

In Eq.~\ref{eq:state1}, we have assumed that Alice produces her high-dimensional entangled states via SPDC. Hence she has knowledge of the subspaces that preserve the photon correlations, and does not need to probe all possible partitions in Fig.~\ref{fig:Figure1}. If Alice had no knowledge of the photon correlation beforehand, how would that affect the computation of the ${\cal P}$-concurrence? In this instance, Alice would need to probe all possible partitions by applying different sets of operators ${\cal B}$. The number of sets $S$ of  operators is given by the number of ways $K$ qubit states of photon B can be matched to $K$ qubit states of photon A: that is $K!$. For $d=3$, there are $S = 6$ sets of  operators ${\cal B}$. Thus, one can produce the 6 distinct diagrams shown in Fig.~\ref{fig:Figure1}(a)-(f). In this instance, the ${\cal P}$-concurrence has the following expression:
\begin{equation}
{\cal P}(\rho_{AB}) = \max_{S}\bigg\{\prod_{k = 1}^{K}{\cal C}(\rho^k_{AB})\bigg\},\label{eq:pconc2}
\end{equation}
where the maximum is taken over all the distinct sets. 

We anticipate that our measure would also be applicable to mixed high-dimensional states since the concurrence equally applies to mixed qubit states. By projecting a high-dimensional mixed state onto lower dimensional spaces, mixed states correlations are preserved. Thus computing the degree high-dimensional entanglement in terms of a qubit measure that is adequate for mixed states would accurately estimate the degree of entangled of the high-dimensional mixed state.
 
In this work, we have introduced a new measure of high-dimensional entanglement beyond the qubit that we call the ${\cal P}$-concurrence. The computation of the ${\cal P}$-concurrence is based on the product of the concurrence over all the qubits subspaces that forms the high-dimensional state. We showed that it is this formulation that ensures that the ${\cal P}$-concurrence estimates the degree of entanglement to a scalar between 0 and 1, More than an entanglement measure, it also acts as a dimension witness given that the ${\cal P}$-concurrence in $d$-dimensions is non-vanishing for any states that does not exhibit entanglement correlations in at least $d$ dimensions. Interestingly, we also show that one does not need to reconstruct the full density matrix through state tomography to compute the ${\cal P}$-concurrence, thus promising faster processing time. 

B.N. acknowledges funding from the National Research Foundation of South Africa.

\end{document}